\documentclass{sf2a-conf2017}
\usepackage{graphicx}
\usepackage{hyperref}
\usepackage[]{natbib}  
\usepackage{epstopdf}

\def\BibTeX{{\rm B\kern-.05em{\sc i\kern-.025em b}\kern-.08em
    T\kern-.1667em\lower.7ex\hbox{E}\kern-.125emX}}
\bibpunct{(}{)}{;}{a}{}{,}  

\begin{document}

\TitreGlobal{SF2A 2017}


\title{Ionization ratios and elemental abundances in the atmosphere of 68 Tauri}

\runningtitle{HD 27962}

\author{A.Aouina}\address{LESIA, UMR 8109, Observatoire de Paris Meudon, Place J.Janssen, Meudon, France}
\author{R.Monier}\address{LESIA, UMR 8109, Observatoire de Paris Meudon, Place J.Janssen, Meudon, France}


\setcounter{page}{237}


\maketitle


\begin{abstract}

We have derived the ionization ratios of twelve elements in the atmosphere of the star 68 Tauri (HD 27962) using an ATLAS9 model atmosphere with 72 layers
computed for the effective temperature and surface gravity of the star.
We then computed a grid of synthetic spectra generated by SYNSPEC49  based on an ATLAS9 model atmosphere in order to model one high resolution spectrum secured by one of us (RM) with the \'echelle spectrograph SOPHIE at Observatoire de Haute Provence.
We could determine the abundances of several elements in their dominant ionization stage, including those defining the Am phenomenon. We thus provide new abundance determinations for 68 Tauri using updated accurate atomic data retrieved from the NIST database which extend previous abundance works.

\end{abstract}



\section{Introduction}

68 Tauri (HD 27962) is the hottest and most massive member of the Hyades open cluster (age about 700 Myrs). Previous abundance analyses of HD 27962 have revealed a distinct underabundance of scandium and overabundances of the iron-peak and heavy elements which prompted to reclassify this early A star as an Am star. The last abundance analysis dates back from 2003 ( Pintado \& Adelman., 2003). It seems therefore justified to redetermine and expand the chemical composition of this interesting object using updated atomic data.

\section{Observations and reduction}

HD 27962 was observed with the SOPHIE spectrograph at the Observatoire de Haute Provence on 04 October 2006 using the High Resolution (R=75000) mode .
The exposure time was 600 seconds yielding a signal-to-noise ratio of about 429 at 5000 \AA. There are currently about 60 high resolution spectra of 68 Tauri in the SOPHIE archive.

\section{Fundamental parameters determination}

In order to derive the effective temperature and surface gravity for 68 Tauri, we
used Napiwotzki et al. (1993) UVBYBETA code
and the observed uvby photometry from Hauck $\&$Mermilliod
(1998) for 68 Tauri. We derived $T_{eff}=9025$ K $\pm 200$ K and $log \ g = 3.25 \pm 0.25$ dex.

\section{Model atmospheres and spectrum synthesis}

Using the stellar parameters derived in the previous section, we have computed a 72 layers plane parallel model atmosphere using the ATLAS9 code (Kurucz, 1992). We justify the use of this model by discussing the assumptions made by ATLAS9 :   
\begin{itemize}
    \item[-] The thickness of atmosphere is  $ 10 ^ {- 3}$ of the stellar radius, so is negligible compared to this radius, which justifies the use of plane-parallel geometry.
    \item[-]  68 Tauri is an Am star so it doesn't have a large magnetic field which could influence the atmospheric structure.
    \item[-] The hypothesis of local thermal equilibrium (LTE) and radiative equilibrium are justified because 68 Tauri is still near the main sequence in the HR diagram.
\end{itemize}

 A plane parallel model atmosphere assuming radiative equilibrium, hydrostatic equilibrium and local thermodynamical equilibrium has been first computed using the ATLAS9 code \citep{Kurucz92}, specifically the linux version using the new ODFs maintained by F. Castelli on her website\footnote{http://www.oact.inaf.it/castelli/}. The linelist was built starting from Kurucz's (1992) gfhyperall.dat file  \footnote{http://kurucz.harvard.edu/linelists/} which includes hyperfine splitting levels.
This first linelist was then upgraded using the NIST Atomic Spectra Database 
\footnote{http://physics.nist.gov/cgi-bin/AtData/linesform} and the VALD database operated at Uppsala University \citep{kupka2000}\footnote{http://vald.astro.uu.se/~vald/php/vald.php}.
A grid of synthetic spectra was then computed with SYNSPEC49 \citep{Hubeny92} to model the lines. The synthetic spectrum was then convolved with a gaussian instrumental profile and a parabolic rotation profile using the routine ROTIN3 provided along with SYNSPEC49.
We adopted a projected apparent rotational velocity $v_{e} \sin i =  10.5 $ km.s$^{-1}$  from \cite{Royer07} and a radial velocity $v_{rad} = 39.43 $ km.s$^{-1}$ determined by cross-corelation of the observed spectrum with the synthetic spectrum.

\section{The ionisation profiles}

In order to determine the abundance of a chemical element, it is preferable to model the absorption lines of the ion corresponding to the  major ionisation state in the atmosphere. To determine the dominant ionisation state for a given element, we have used Saha's equation which yields the fraction between the number of ions in two successive ionisation states versus optical depth, assuming  LTE in the atmosphere of the star :
$$\frac{n_j(\tau)}{n_{j-1}(\tau)}=\frac{\phi_j(T)}{P_e}$$ 
      where: $$\phi_j(T)=0.665\frac{u_j}{u_j-1}T(\tau)^{\frac{5}{2}}10^{\frac{-5040I}{kT(\tau)}} $$
Using the temperature profile  $T(\tau)$ from the model atmosphere, the partition function $u$ of each ion and the ionisation potential $I$,  we could derive the ionisation profiles of various elements.\\ 
   
\begin{figure}

    \centering
    \includegraphics[width=0.7\linewidth]{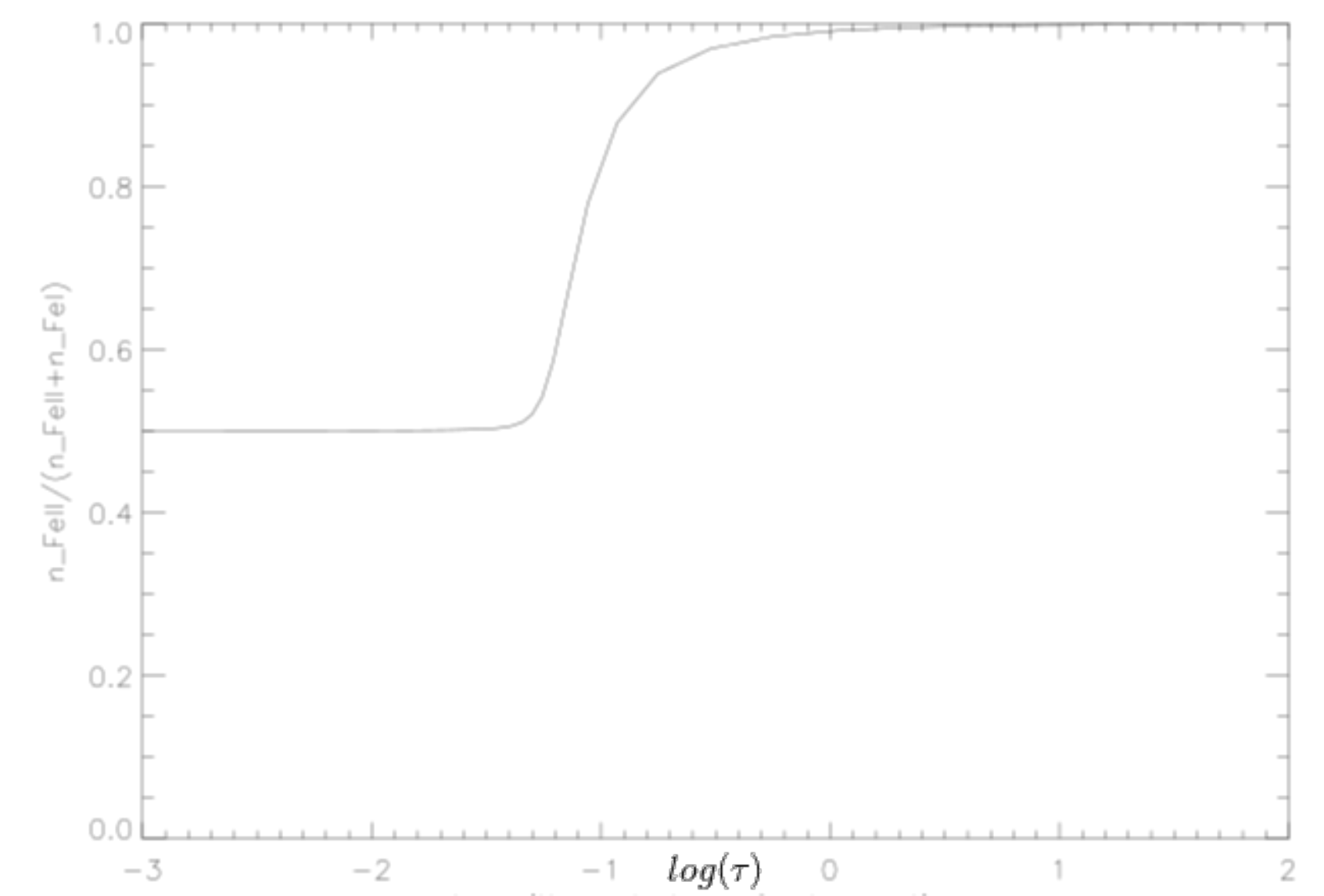}
    \caption{Ionisation ratio of Fe II versus  optical  depth $\tau$ }
    \label{fig1}
\end{figure}

Fig.~\ref{fig1} represents the run of the ionisation ratio of once ionised iron FeII over the total number of neutral and ionised iron versus optical  depth in the atmosphere of 68 Tauri. We can see that FeII is the dominant ionization stage, which justifies the use of the Fe II lines rather than FeI to derive LTE abundances of iron in the atmosphere of 68 Tauri.

\section{Abundances determination method}

To derive the abundances of the various chemical elements in the atmosphere of 68 Tauri, we have iteratively adjusted synthetic spectra to selected observed line profiles. It In SYNSPEC49 the theoretical flux in LTE is computed as:

$$F_{\nu}(0,\mu)=2 \pi \int_{0}^{\infty} I_{\nu} (0,\mu) \mu d\mu $$

where $ I_{\nu} $ is the intensity of the radiation field from and $ \mu $ the cosine of the angle between the direction of propagation and the area traversed.
\\ This flux is a complex function of several parameters: 
$$F=F_\nu(T(\tau_{\nu}),\xi,log \ gf, \gamma_i, \frac{X}{X_\odot})$$

where $T(\tau_{\nu})$ is the temperature at each optical depth, $\xi$ the microturbulent velocity assumed to be constant with depth, $gf$ the product of the statistical weight and the oscillator strength, $\gamma_i $ the damping constants and $ \frac{X}{X_\odot} $ the relative abundance compared to the solar value.

We have fixed as many physical parameters as possible and left the unknown abundance as the only free parameter. Then for each line of each element we changed the abundance until the synthetic spectrum fits best the observed spectrum normalized to its local continuum.
A priori, we analysed unblended lines whose atomic parameters are as accurate as possible.

\section{Results and discussion }

The abundance analysis of 68 Tauri yields a distinct under-abundance of Sc and over-abundances of  V, Cr, Ni, Dy and Ba. These anomalies are characteristic features of hot Am stars.
Pintado and Adelman (2003) obtained the abundances of 32 elements in 68 Tau. They used the same atmospheric parameters as in this work. Their
results agree with ours within $\pm 0.25 dex$ for Mg, Ni, Sc, Ca, Fr, Ti, and Ba. However,  the abundances of vanadium disagree because of the difference in the $log \ gf$ value of V II adopted in this work  and in Pintado $\&$ Adelman (2003). Indeed we have used more recent improved $log \ gf$ value from the NIST database.
      \begin{table}
          \centering
          \begin{tabular}{l l l l}
           \hline \hline \centering
   \ \ \ \ \ & O$\&$S (1990)& P$\&$A (2003) & This work   \\ 
    \hline \hline
     Fe & 0.81 $\odot$ & 1.51$\odot$ & 2.18$\odot$
      \\ Ti&  1.31$\odot$ &  1.77$\odot$& 2.68$\odot$
      \\Cr& 4.67$\odot$& 3.98$\odot$ & 3.41$\odot$
      \\ Mg & 1.41$\odot$&1.62$\odot$& 1.50$\odot$
     \\ Ca &  0.34$\odot$& 0.88$\odot$ & 0.70$\odot$
     \\ V &  5.37$\odot$& 0.69$\odot$& 4.42$\odot$
     \\ Ba &  9.33$\odot$& 21.87$\odot$ & 23.00$\odot$ 
    \\ Ni & 0.75$\odot$ & 8.91$\odot$ & 8.50$\odot$
    \\ Sc & 0.08$\odot$& 0.12$\odot$ & 0.06$\odot$
        \\ Dy & $\times$ & 2.01$\odot$ & 1.60$\odot$
    \\ Sm & $\times$& 2.10$\odot$ & 2.40$\odot$
    \\ Ce & $\times$& 1.90$\odot$ & 1.71$\odot$
    \\ \hline
        \end{tabular}
          \caption{Comparison of the derived abundances with previous works}
          \label{tab: compa}
      \end{table}

\section{Conclusions}

Our abundance analysis using recent atomic data confirms the Am status of HD 27962 and provides improved abundances for Fe, Ti, Cr, Mg, Ca, Sc, V, Ni, Ba, Dy, Sm and Ce.
A continuation of this work will consist in using all spectra available in the SOPHIE archive to derive more elemental abundances and search for radial velocity and line variations. 

\begin{acknowledgements}
The authors acknowledge use of the SOPHIE archive (\url{http://atlas.obs-hp.fr/sophie/}) at Observatoire de Haute Provence. They have used the NIST Atomic Spectra Database and the VALD database operated at Uppsala University (Kupka et al., 2000) to upgrade atomic data.
\end{acknowledgements}



\begin{thebibliography}{}

\bibitem[Hubeny \& Lanz (1992)]{Hubeny92} Hubeny, I., Lanz, T. 1992, A\&A, 262, 501
\bibitem[Kupka et al. (2000)]{kupka2000}Kupka F., Ryabchikova T.A., Piskunov N.E., Stempels H.C., Weiss W.W.,
  2000, Baltic Astronomy, vol. 9, 590-594
\bibitem[Kurucz (1992)]{Kurucz92} Kurucz, R.L. 1992, Rev. Mexicana. Astron. Astrofis., 23, 45
\bibitem[Monier et al. (2015)]{Monier15} Monier, R., Gebran, M., Royer, F. 2015, A\&A, 577A, 96M
\bibitem[Monier et al.(2016)]{Monier2016} Monier, R., Gebran, M., \& Royer, F.\ 2016, \apss, 361, 139 
\bibitem[Napiwotzki et al. (1993)]{Napiwotzki93} Napiwotzki, R., Schoenberner, D., Wenske, V. 1993, A\&A, 268, 653
\bibitem[Okyudo \& Sadakane (1990)]{okyudo90} Okyudo, M., Sadakane, K., 1990, PASJ, 42, 317O
\bibitem[Adelman et al. (2003)]{adelman03} Pintado, O. I.; Adelman, S. J., 2003, A\&A, 406, 987
\bibitem[Royer et al. (2007)]{Royer07} Royer, F., Zorec, J., Gomez, A.E. 2007 , A\&A, 463, 671R
\end{thebibliography}

%
\end{document}